\documentclass[12pt]{raa}
\usepackage{graphicx,times}
\usepackage{natbib}
\usepackage{epsfig}
\usepackage{amsmath}
\providecommand{\bm}[1]{\mbox{\boldmath$#1$\unboldmath}}

\newcommand{\chandra}{{\it Chandra}}

\newcommand{\ASCA}{{\it ASCA}}

\begin{document}

\title{\chandra\ Study of the Massive and Distant Galaxy Cluster SDSS J0150-1005 }

\author{Zhenzhen Qin \inst{1}
  \and Haiguang Xu\inst{1}
  \and Jingying Wang\inst{1}
  \and Junhua Gu\inst{2}
}

\institute{
$^{1}$Department of Physics, Shanghai Jiao Tong University,
800 Dongchuan Road, Shanghai 200240, China; 
{\it qzz@sjtu.edu.cn}\\
$^{2}$National Astronomical Observatories, Chinese Academy of Sciences, 20A Datun Road, Beijing 100012, PRC}

\abstract{
In this work, we present a high spatial resolution study of a fossil cluster, SDSS J0150-1005 ($z \simeq 0.364$), based on the imaging spectroscopic analysis of \chandra\ observation. The \chandra\ X-ray image shows a relax and symmetric morphology, which indicates SDSS J0150-1005 is a well-developed galaxy cluster with no sign of recent merger. Its global gas temperature is 5.73 $\pm$ 0.80 keV, and the virial mass is 6.23 $\pm$ 1.34 $\times 10^{14} $ M$_{\odot}$  according to the isothermal model. Compared with polytropic temperature model, the mass calculated based on isothermal model overestimates 49 $\pm$ 11\%. The central gas entropy, $S_{0.1r_{200}} = 143.9 \pm 18.3$ keV cm$^2$, is significantly lower than average value of the normal galaxy clusters with similar temperatures. Our results indicate the early formation epoch of SDSS J0150-1005.
}

\authorrunning{Zhenzhen Qin et al.}
\titlerunning{Fossil Cluster SDSS J0150-1005}
\maketitle

\section{INTRODUCTION}
Galaxy clusters are the largest gravitational bound systems in the universe, which originated from extreme overdensities in the primordial density field, and grew through continuous accretion as well as serial mergers, and finally formed mass concentrations of $10^{14-15}$ M$_{\odot}$ at the present epoch. X-ray study of galaxy cluster plays an important role in establishing the current cosmological paradigm. However, it is still necessary to investigate massive and distant galaxy clusters to help determine the cosmological parameters with even higher precision.

SDSS J0150-1005 (RA=01h50m21.3s, Dec=$-$10d05m31s, J2000.0) is located at $z \simeq 0.364$. It is a massive (and thus, by inference, very X-ray luminous), distant galaxy cluster in the MACS (the MAssive Cluster Survey) sample, which includes the most extreme and rarest clusters out to significant redshift (Ebeling et al. 2001). SDSS J0150-1005 is identified as a fossil system by Santos et al. (2007), in which 34 fossil systems are found from the Sloan Digital Sky Survey (Adelman-McCarthy et al. 2007). According to the definition of Jones et al. (2003), fossil system is a gravitational bound system with spatially extended X-ray emission, whose X-ray luminosity is higher than $1 \times 10^{42}$ ergs $\rm s^{-1}$ and $\Delta m_{\rm 12} \geq 2.0$ mag, where $\Delta m_{\rm 12}$ is the absolute total magnitude gap in R band between the brightest and second-brightest member galaxies within half of the (projected) virial radius. The origin and evolution process of the fossil systems are still not well understood.  In this work, we study a galaxy cluster, SDSS J0150-1005, by analysing \chandra\ archive data to probe the X-ray properties of fossil system in the very high mass regime.

We describe the data reduction, imaging and spectroscopy analysis in ${\S}$ 2, discuss and summarize our results in ${\S}$ 3 and 4, respectively. Throughout this paper, we adopt the angular size distance of 1031.7 Mpc to the cluster using $H_{0}=71$ km s$^{-1}$ Mpc$^{-1}$,
$\Omega_{m} = 0.3$ and $\Omega_{\Lambda} = 0.7$. At this distance, $1^{\prime}$  corresponds to 296.9 kpc. We adopt the solar abundance standards of Grevesse and Sauval (1998), where the iron abundance relative to hydrogen
is $3.16\times10^{-5}$ in number. Unless stated otherwise, the quoted errors are the 68\% confidence limits.

\section{DATA ANALYSIS}
\label{sect:data}

\subsection{Observation and data reduction}
The \chandra\ observation of SDSS J0150-1005 was carried out on September 14, 2009 (ObsID 11711)
for a total exposure of 27.1 ks with CCDs 0, 1, 2 and 3 of the Advanced  CCD Imaging Spectrometer (ACIS) in operation. The events were collected  with a frame
time of 3.2 s and telemetered in the VeryFaint mode. The focal plane temperature was set to -120 $^{\circ}$C. We use the \chandra\ data analysis package CIAO software (version 4.5) to process the data extracted from the I chips. We keep events with \ASCA\ grades 0, 2, 3, 4 and 6, and remove all the bad pixels, bad columns, columns adjacent to bad columns and node boundaries. In order to identify possible strong background flare, light curves are extracted from a box region $5.1^{\prime}$ (about $1.5r_{500}$) far away from the X-ray peak. Time intervals during the count rate exceeding the average quiescent value by 20\% are excluded.

\subsection{X-ray surface brightness}

Fig. 1a shows the raw \chandra\  ACIS-I image of SDSS J0150-1005 in $0.7-7.0$ keV band, where all the point sources of the fossil cluster that could be detected at the confidence level of $3\sigma$ by the CIAO tool ${celldetect}$ were excluded. The X-ray morphology roughly shows a relax and symmetry appearance. Fig. 1b shows the SDSS B-band image for the box region in Fig. 1a, with the Chandra $0.7-7.0$ keV X-ray intensity contours overlaid. The white cross in the center indicates the X-ray peak of SDSS J0150-1005. The X-ray peak and the position of the cD galaxy SDSS J015021.27-100530.4 are consistent with each other within $1^{\prime\prime}$. The cD galaxy shows the regular morphology, and presents as a giant isolated elliptical galaxy.

   \begin{figure}[h!!!]
   \centering
   \includegraphics[width=14.0cm, angle=0]{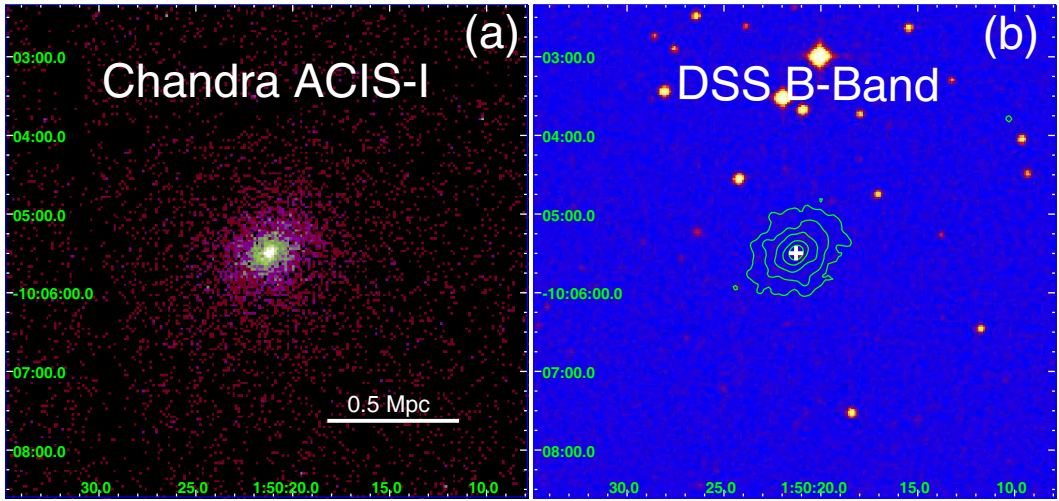}
   \caption{
     (a): Raw \chandra\ image of SDSS J0150-1005 in the $0.7-7.0$ keV energy band.
     (b): SDSS B-band optical image for the box region in Fig. 1a, where the X-ray contours of \chandra\ $0.7-7.0$ keV image are overlaid. The white cross in the center indicates the X-ray peak of SDSS J0150-1005.
   } 
   \label{Fig1}
   \end{figure}

The exposure-corrected surface brightness profile (SBP) is extracted from a series of 
annuli regions centered on the X-ray emission peak, and shown in Fig. 2. The energy band of SBP is restricted to the $0.7-7.0$ keV. Assuming the spherical symmetry, we deproject the SBP to derive the 3-dimensional electron number density profile $n_{\rm e}(r)$ of the ICM with standard ``$onion-skin$'' method (Kriss et al. 1983). Using the single-$\beta$ model, we fit the $n_{\rm e}(r)$ of SDSS 0150-1005 as:

\begin{equation}
n_{\rm e}(r) = n_{\rm 0}[1+(r/r_{\rm c})^2]^{-1.5\beta}+n_{\rm bkg},
\end{equation}
where $r$ is the radius, $n_{\rm 0}$ corresponds to the central electron number density, $r_{\rm c}$ is the core radius, $\beta$ is the slope and $n_{\rm bkg}$ is the background. The reduced chi-square of the single-$\beta$ model fit is $\chi^2/dof = 1.88$, which is unsatisfactory. The single-$\beta$ best-fit  SBP (dash line in Fig. 2) exhibits significant disagreement in the central 30 kpc due to the existence of central excess. Therefore, we apply the double-$\beta$ model to describe the $n_{\rm e}(r)$ as,
\begin{equation}
n_{\rm e}(r) = n_{\rm 1}[1+(r/r_{\rm c1})^2]^{-1.5\beta_1}+n_{\rm 2}[1+(r/r_{\rm c2})^2]^{-1.5\beta_2}+n_{\rm bkg},
\end{equation}
which gives acceptable fit with the reduced chi-square $\chi^2/dof = 1.04$. The double-$\beta$ best-fit SBP is shown as solid line in Fig. 2. The resultant $n_{\rm 1} = 0.054 \pm 0.004$ cm$^{-3}$, $r_{\rm c1} = 15.48 \pm 1.59$ kpc, $\beta_1 = 0.65 \pm 0.02$ for one component; $n_{\rm 2} = 0.005 \pm 0.001$ cm$^{-3}$, $r_{\rm c2} = 107.43 \pm 14.75$ kpc, $\beta_2 = 0.64 \pm 0.02$ for another. According to the best-fit gas distribution, the average gas density in the central 150 kpc region is  $\simeq$ 0.005 $\pm$ 0.001 cm$^{-3}$.

   \begin{figure}[h!!!]
    \centering
     \includegraphics[angle=0, width=9.0cm]{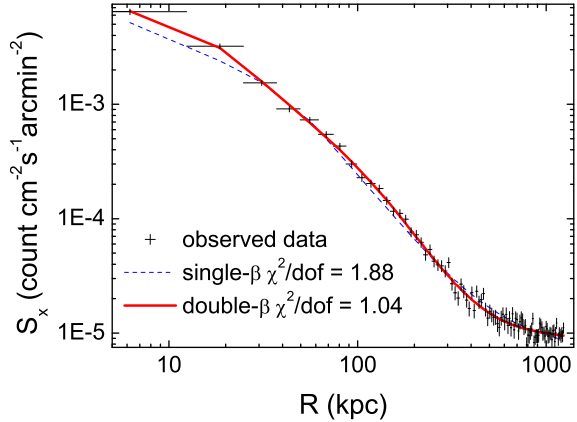}
     \caption{Exposure-corrected surface brightness profiles extracted from a series of annuli regions centered on the X-ray emission peak of SDSS J0150-1005. R is the projected distance. The blue dash and red solid lines correspond to the best-fit single-$\beta$ and double- $\beta$ SBPs, respectively.
       }
      \end{figure}

\subsection{Hardness ratio map}
We define the hardness ratio as the ratio of counts within $2.0-7.0$ keV band over that within $0.7-2.0$ keV band. For the ICM observations, the hardness ratio is a temperature diagnostic: larger hardness ratio means higher temperature. The hard-band ($2.0-7.0$ keV) and the soft-band ($0.7-2.0$ keV) counts are extracted from the events data, and then binned spatially by a factor of 20 to increase the statistical significance for every pixel. Fig. 3 exhibits the hardness ratio map, which clarifies the SDSS J0150-1005 may be a cool core cluster, because it has less hard-band count in the center. Moreover, there is no obvious thermal substructure in this cluster.

   \begin{figure}[h!!!]
    \centering
     \includegraphics[angle=0, width=9.0cm]{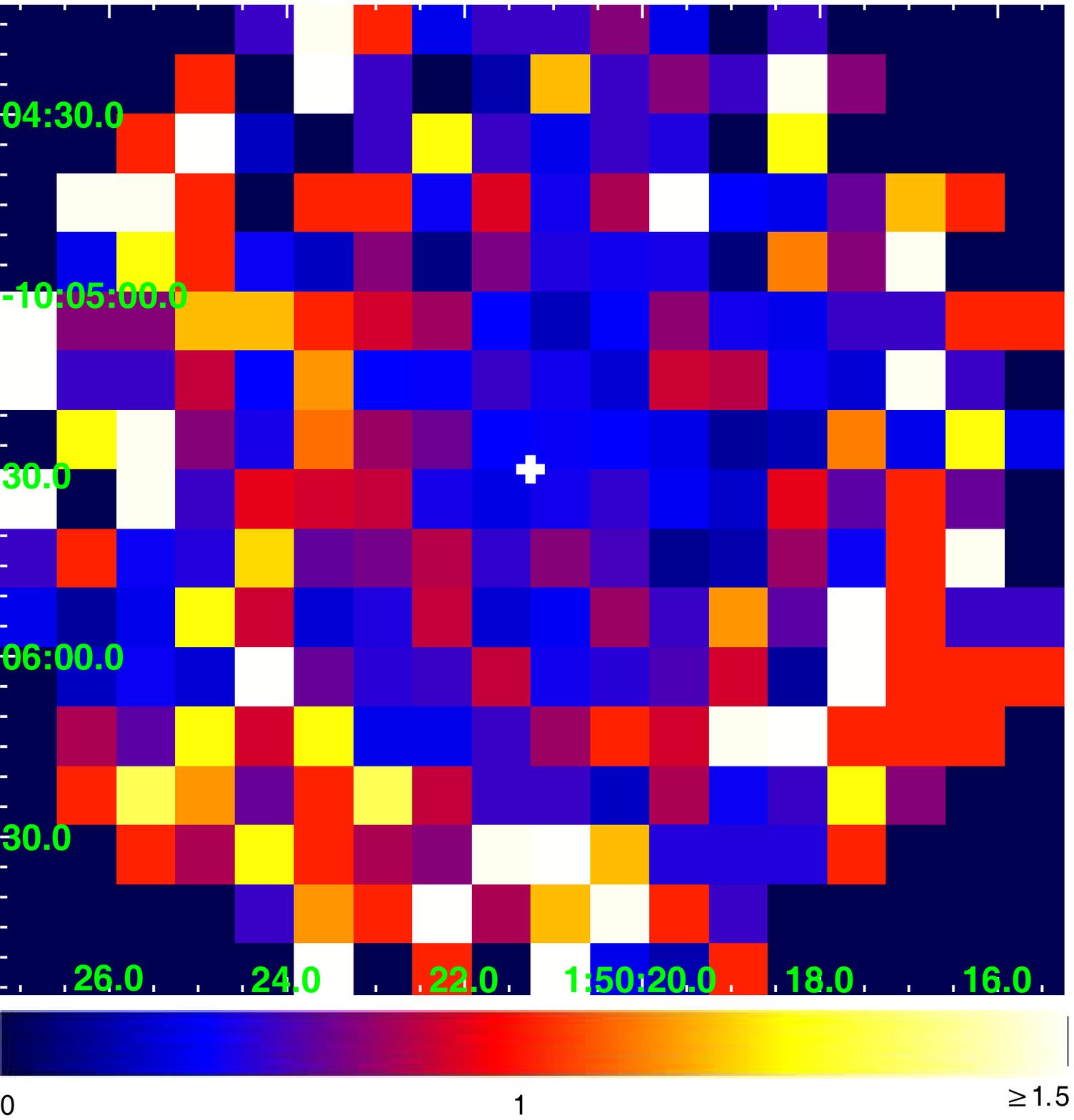}
     \caption{Hardness ratio map (see definition in ${\S}$ 2.3), which is binned spatially by a factor of 20. The white cross in the center indicates the X-ray peak of SDSS J0150-1005.
       }
      \end{figure}

\subsection{Spectral analysis}

 We utilize the \chandra\ blank-sky template for the ACIS CCDs as the background. The template is tailored to match the actual pointing (Pointing RA = 27.5947774587, Pointing Dec = -10.0645622317 and Pointing Roll = 81.4228143940). The background spectrum is extracted and processed identically to the source spectrum. Then, we rescale the background spectrum by normalizing its high energy end $9.0-12.0$ keV to the corresponding observed spectrum. The corresponding spectral redistribution matrix file (RMFS) and auxiliary response files (ARFS) are created by using the CIAO tool $mkacisrmf$ and $mkwarf$, respectively. All spectra are rebinned to ensure at least 20 raw counts per spectral bin to allow $\chi^2$ statistics to be applied. Since the contribution of the hard spectral component is expected to be rather weak, and also to minimize the effects of the instrumental background at higher energies as well as the calibration uncertainties at lower energies, the fit is restricted to the $0.7-7.0$ keV. 

Due to the limited counts, it is insufficient to extract the deprojected temperature profile of SDSS J0150-1005. We take the spectrum temperature from annulus of $0.2-0.5r_{500}$ as the global temperature in order to reduce the effect of cool core (Zhang et al. 2007). $r_{500}$ is the radius, within which the average mass density is 500 times of the critical density of the universe at corresponding redshift. We will describe how to determine it in  ${\S}$ 3.1. We use the XSPEC 12.4.0 package to fit the spectrum. Following the method in Gu et al. (2010),  we model the gas with WABS $\times$ APEC model. We fix the absorption to the Galactic value of 2.15 $\times 10 ^{20}$ cm$^{-2}$ (Dickey \& Lockman 1990). The best fit determines a temperature of 5.73 $\pm$ 0.80 keV , and an abundance of 0.30 $\pm$ 0.07 times solar with $\chi^2/dof = 0.97$. As a cross check for its cool core identified with the hardness ratio map, we also extract the spectral temperature within the $0.2r_{500}$ and fit the spectrum with the same method described above. In the cluster center, the best-fit ($\chi^2/dof = 1.02$) temperature and abundance are $4.54\pm0.33$ keV and 0.51 $\pm$ 0.14 times solar, respectively. Such low inner temperature also indicates the cool core of SDSS J0150-1005.

\section{Results}


\subsection{Total gravitational mass profile}
In a spherically symmetric system with hydrostatic equilibrium, $M_{\rm tot}(<r)$, the total mass within a given radius, $r$, is given by
\begin{equation}\label{eq.m}
M_{\rm tot}(<r) = -{{k_{\rm B} T(r) r} \over {G \mu m_{\rm p}}}[{{d\  {\rm ln} n_{\rm e}(r) } \over {d \ {\rm ln} r}} +{{d \ {\rm ln} T(r)} \over {d \ {\rm ln} r}}], 
\end{equation}
where $G$ is the universal gravitational constant, $k_{\rm B}$ is the Boltzmann constant, $\mu=0.62$ is the mean molecular weight per hydrogen atom, $m_{\rm p}$ is the proton mass, $n_{\rm e}(r)$ is the electron number profile, and $T(r)$ is the temperature profile. In the $M_{\rm tot}(<r)$ calculation, $n_{\rm e}$ is obtained from the best-fitting of the X-ray surface brightness profile described in ${\S}$ 2.2. As described in ${\S}$ 2.4, we can not get the deprojected temperature profile of SDSS J0150-1005. Therefore, we assume this cluster is an isothermal system. In such case, there is an iterative process involved in the mass calculation: an initial $r_{500}$ is provided to obtained the global temperature; the global temperature are then used in Eq. (3) to produce $r_{500}$ for the next iteration; such iteration continues until $r_{500}$ and global temperature converges. Based on the resultant global temperature, the virial radius $r_{200}$ and virial mass $M_{200}$ are obtained. With Eq. (\ref{eq.m}) and 1000 Monte-Carlo simulations, virial radius $r_{200}$ = 1.54 $\pm$ 0.11 Mpc , and the virial mass is 6.23 $\pm$ 1.34 $\times 10^{14} $ M$_{\odot}$. The total gravitational mass profile is shown in Fig. 4.

Assuming isothermal distribution may overestimate the mass of galaxy cluster. In order to constrain this uncertainty, we also adopt the polytropic model (Sarazin 1988), where temperature gradient in the intracluster gas is assumed.  In this model, if the hot gas is strictly adiabatic, its pressure and density have a simple relation of $P \propto \rho_{gas}^{\gamma}$, polytropic index $\gamma$ is the usual ratio of specific heat. Because of the low density ($n_e < 10^{-2}$ cm$^{-2}$), the ICM can be treated as an ideal gas with $P = n_{gas}k_BT$. The temperature profile is described by 
\begin{equation}\label{eqTprofile}
T(r) = T_{0,\gamma}({{\rho_{gas}(r)} \over {\rho_0}})^{\gamma-1} = T_{0,\gamma}({{n_{e}(r)} \over {n_0}})^{\gamma-1}.
\end{equation}
As a rough estimation, we adopt the polytropic index $\gamma = 1.19 \pm 0.03$ according to Tawa (2008), in which they used Suzaku observations and extended the temperature profiles up to the virial radii. The resultant virial mass is 4.20 $\pm$ 1.01 $\times 10^{14} $ M$_{\odot}$, and the calculated total gravitational mass profile is also shown in Fig. 4. We find the virial mass derived from isothermal model is overestimated by 49 $\pm$ 11\% compared with polytropic temperature model.

  \begin{figure}[h!!!]
    \centering
    \includegraphics[angle=0,width=9.0cm]{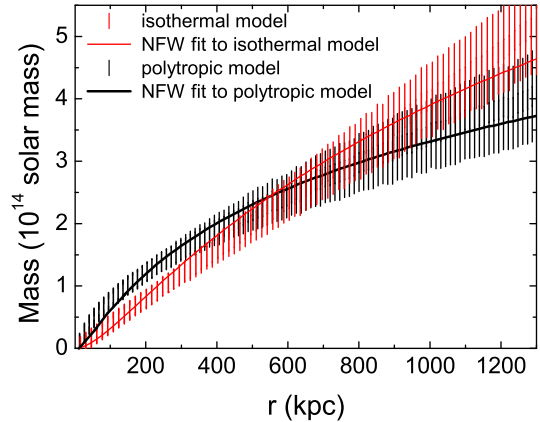}
       \caption{Total gravitational mass profiles of SDSS J0150-1005 based on isothermal (red) and polytropic (black) temperature models. The solid lines present the best fits to the NFW model, which give mass concentration parameters $c_{200} = 8.55 \pm 0.38$ and 22.46 $\pm$ 1.33 for the isothermal and polytropic temperature models, respectively.}
  \end{figure}




  
\subsection{Gas luminosity}
   The luminosity, $L_{\rm X}$, of SDSS J0150-1005 within $r_{200}$ is given by $L_{\rm X}=\iiint \Lambda n_{\rm e} n_{\rm H} dV$, where $\Lambda$ is cooling function and $n_{\rm H}$ is the proton number density. We calculate the integral of electron number density profile to get the bolometric $L_{\rm X} = 3.87 \pm 1.47 \times 10^{44}$ erg s$^{-1}$. The error of $L_{\rm X}$ is obtained from the Poisson error in the X-ray count rate.

\section{DISCUSSION}
According to results presented in ${\S}$ 3, we would like to discuss the formation epoch of SDSS J0150-1005 through X-ray morphology, mass concentration parameter and central entropy as follows.

The X-ray morphology of SDSS J0150-1005 shows a relax and symmetry appearance: the cD galaxy is consistent with the X-ray peak within $1^{\prime\prime}$; there exists a cool core in the center of galaxy cluster as shown in the hardness ratio map; and also no clumps or substructure is seen in the ICM. According to the simulation of Poole et al. (2006), which claims that the system needs $\sim 2 \times 10^9$ yr to relax and virialize, and the smoothness of its X-ray surface brightness isophotes is the only reliable indicator of the virialization. Therefore, this galaxy cluster exhibits no violently turbulence caused by recent major mergers.

We apply the NFW profile (Navarro et al. 1995)
\begin{equation}
\rho(r) = {\rho_{\rm 0} \delta_{\rm c} \over (r/r_{\rm s})(1+r/r_{\rm s})^2},
\end{equation}
to describe the total gravitational mass density, where $\rho$ is the mass density, $r_{\rm s}$ is the scale radius, $\rho_{\rm 0}$ is the
critical density of the universe, and $\delta_{\rm c}$ is the characteristic density. With the NFW model, we can obtain the following integrated profile of a spherical
mass distribution,
\begin{equation}\label{eq_m}
M_{\rm tot}(<r) = 4 \pi \delta_{\rm c} \rho_{\rm 0} r_{\rm s}^3[\ln(1+{r \over r_{\rm s}}) - {r \over r+r_s}].
\end{equation}
We fit the NFW mass profile, Eq. (\ref{eq_m}), to the X-ray derived isothermal-model total gravitational mass profile obtained in ${\S}$ 3.1, and the result is shown in Fig. 4. The best-fit gives  $r_{\rm s}$ = 181.24 $\pm$ 11.20 kpc. The mass concentration parameter is defined as $c_{200} = r_{200}/r_{\rm s} = 8.55 \pm 0.38$. Due to the systematic overestimation of the mass distribution by the isothermal model, we also apply the NFW mass profile to the total gravitational mass profile derived by polytropic temperature model in ${\S}$ 3.1, the resultant  $r_{\rm s}$ =60.48 $\pm$ 5.86 kpc and $c_{200} = 22.46 \pm 1.33$. Both concentration parameters are higher than the numerical and observed values in galaxy clusters of similar mass (Dolag et al. 2004; Pratt \& Arnaud 2005). Simulation shows that, dark matter halos, which have not undergone a major merger since $z = 2$, are more concentrated than those have experienced one since then (Wechsler et al. 2002). Such a high mass concentration may indicate an early forming epoch.

The gas entropy is defined as
\begin{equation}
S= T/n_{\rm e}^{2/3}.
\end{equation}
We focus on the gas entropy at $0.1r_{\rm 200}$, $S_{ 0.1r_{200}}$. Because $0.1r_{200}$ is very close to the center, so that we can avoid the shock-generated entropy, and thus enhance the sensitivity to any additional entropy. This yields $S_{0.1r_{200}} = 143.9 \pm 18.3$ keV cm$^2$, which is significantly lower than average value obtained by Pratt et al. (2006) for the relax galaxy clusters with similar temperatures. Voit et al. (2003) suggested that most feedback within the ICM operates before it falls into the clusters, other than after virialization. In the pre-collapse gas,  the high density contrast leads to higher post-shock entropy, once the gas crosses the accretion shock during the galaxy cluster formation. According to above statements, we conclude that the lower central gas entropy also indicates the early forming epoch of this galaxy cluster.

As discussed above, SDSS J0150-1005 shows no recent major merger and early formation according to its relaxed X-ray morphology, high mass concentration and the low central entropy. From the simulation by D'Onghia et al. (2005), a fossil system should have already assembled half of their final mass at z $\sim$ 1, and subsequently they typically grow by minor merging only. The high mass concentration was also reported in other fossil systems previously (Khosroshahi et al. 2004; 2006; Sun et al. 2004). Moreover, the gas entropy in fossil systems is reported lie along the lower envelope of the entropy-temperature distribution (Khosroshahi et al. 2007). Therefore, we conclude that the X-ray properties of SDSS J0150-1005 may be ascribed to its fossil character.

\section{SUMMARY}
\label{sect:conclusion}
With the \chandra\ observation of a massive and distant galaxy cluster, SDSS J0150-1005, we study hot-gas properties of this fossil cluster, including its temperature, luminosity and gas entropy. The isothermal and polytropic temperature models are adopted to estimate total gravitational mass. Its relax X-ray morphology, high mass concentration parameter and central low entropy indicate that, this cluster is early formed, and has no recent major merger. We conclude that these X-ray properties may be ascribed to its fossil character.

\normalem
\begin{acknowledgements}
 We thank the anonymous referee for useful suggestions on the manuscript. This work was supported by the Ministry of Science and Technology of the People's Republic of China (973 Program; Grant Nos. 2009CB824900 and 2009CB824904), the National Science Foundation of China (Grant Nos. 10878001, 10973010, and 11125313), and the Shanghai Science and Technology Commission (Program of Shanghai Subject Chief Scientist; Grant Nos. 12XD1406200 and 11DZ2260700).
\end{acknowledgements}

\clearpage


\label{lastpage}

\end{document}